\documentclass[
reprint,
showpacs,
nobibnotes,
superscriptaddress,
longbibliography,
amsmath,amssymb,
aip
]{revtex4-2}

\usepackage[skins,theorems]{tcolorbox}
\usepackage{color}
\usepackage{centernot}
\usepackage{graphicx} 
\graphicspath{{./figures/}}
\usepackage{dcolumn} 
\usepackage{siunitx}
\usepackage{bm} 
\usepackage{bbold}
\usepackage{braket}
\usepackage{hyperref} 
\usepackage[mathlines]{lineno} 
\usepackage{subfigure}
\usepackage{mathtools} 
\usepackage{mathrsfs} 

\renewcommand*\vec{\bm}

\newcommand*\img{\mathop{}\!\text{i}}
\newcommand*\dd{\mathop{}\!\text{d}}

\newcommand*\diff[2]{\mathop{}\!\frac{\dd #1}{\dd #2}}

\begin{document}
\title{Quantum Imaging as alternative readout in atom interferometry: The case of Gravity}

\author{Marian Cepok}
\affiliation{ZARM, University of Bremen, 28359 Bremen, Germany}

\author{Dennis R\"atzel}
\affiliation{ZARM, University of Bremen, 28359 Bremen, Germany}
\affiliation{Gauss-Olbers Space Technology Transfer Center, University of Bremen, 28359 Bremen, Germany}

\author{Claus L\"ammerzahl}
\affiliation{ZARM, University of Bremen, 28359 Bremen, Germany}
\affiliation{Gauss-Olbers Space Technology Transfer Center, University of Bremen, 28359 Bremen, Germany}

\begin{abstract}
We propose a quantum imaging-inspired setup for measuring gravitational fields using an atom that emits a photon at one of two possible locations.
The atom acquires a gravitationally induced quantum phase that it shares with the photon.
By restoring the path identity of the atom after its interaction with the gravitational field, the gravitationally induced phase can be measured using photon interferometry without the need for additional measurements on the atom.
Through repeated measurements with varying interferometric setups, the gravitational potential and inertial acceleration can be deduced.
\end{abstract}

\maketitle


\section{\label{sec:introduction}Introduction}

Since its first emergence, atom interferometry has proven to be a promising instrument for measuring the gravitational field \cite{kasevichMeasurementGravitationalAcceleration1992, petersMeasurementGravitationalAcceleration1999, menoretGravityMeasurements102018}.
Today, it is recognized as one of the most precise methods not only for gravimetry \cite{huDemonstrationUltrahighsensitivityAtominterferometry2013, biedermannTestingGravityColdatom2015}, but also for gradiometry \cite{delaguilaBraggGravitygradiometerUsing2018}, and it is already being used for tests of the equivalence principle \cite{schlippertQuantumTestUniversality2014, taralloTestEinsteinEquivalence2014, asenbaumAtomInterferometricTestofEquivlancePrinciple2020} and local Lorentz invariance \cite{mullerAtomInterferometryTestsIsotropy2008}.
Furthermore, there are plans to detect gravitational waves using atom interferometers \cite{dimopoulosAtomicGravitationalWave2008}.

The indistinguishability of quantum particles also plays a central role in other types of interferometers.
A notable early example is the photon-based Hong-Ou-Mandel interferometer, which demonstrates the effect of indistinguishability in quantum measurements \cite{hongMeasurementSubpicosecondTime1987}, which has since been validated for atoms as well \cite{lopesAtomicHongOu2015}.
Shortly after the Hong-Ou-Mandel experiment, another interferometric setup based on the indistinguishability of photons created in a down-conversion process was proposed by Zou, Wang, and Mandel \cite{zouInducedCoherenceIndistinguishability1991}.
In this setup, which we will refer to as ``Quantum Imaging'' throughout this article and which is used, for example, in\cite{lemosQuantumImagingUndetected2014, gilabertebassetPerspectivesApplicationsQuantum2019}, an object is imaged using a pair of entangled photons, one of which interacts with the object while the other is measured.
Notably, the photons can have different properties, such as belonging to different wavelength ranges.
This is possible because the photon pair forms a composite state with a shared quantum phase.
The phase that one photon acquires through its interaction with the object is transferred to the other photon through a method known as \textit{path identity} \cite{lahiriManyparticleInterferometryEntanglement2018, hochrainerQuantumIndistinguishabilityPath2022}, which involves recombining superimposed paths of one of the photons.

In this paper, we introduce a novel setup that combines optical and matter wave interferometry in which an atom emits a photon at one of two possible locations (see Figure \ref{fig:overview}).
In contrast to conventional Quantum Imaging, this is achieved by replacing one of the photons on a split path with an atom, where the phase accumulated by the atom is a common phase to the composite atom-photon state.
By recombining the atomic paths at the end of the interferometer, the previously accumulated phase is attributed to the coherent superposition, which is now assumed by the photon.
The entire phase can then be read off from the photonic measurement, which allows conclusions to be drawn about the gravitational field.
This may also be viewed as an alternative scheme for the readout of matter wave phases in state-of-the-art matter wave interferometry.
Furthermore, the setup is also interesting for studying the fundamental interface between quantum mechanics and gravitation, because it involves entanglement between two completely different kinds of particles.%

In order to focus on the essentials, we reduce the setup described in this article to an idealized measurement scheme and discuss ideas for practical implementation further below in the article.

\section{\label{sec:schematic}Setup}

\begin{figure}[b]
    \begin{center}
        \includegraphics[width=0.8\columnwidth]{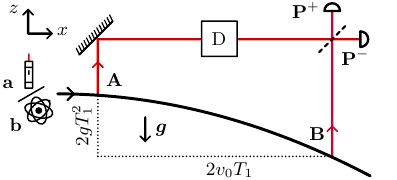}
    \end{center}
    \caption{
        Depicted is an overview of two possible interferometric setups.
        The setups are denoted as \textbf{a} and \textbf{b}.\newline
        \textbf{a.} A classical particle (solid black line) emitting radiation (red lines) is falling in a homogeneous gravitational field.
        The radiation emitted at \textbf{A} and at \textbf{B} is recombined on a beam splitter and measured by photodetectors labeled $\mathbf{P^-}$ and $\mathbf{P^+}$, while no further measurements are performed on the particle.\newline
        \textbf{b.} An atom (solid black line) is falling in a homogeneous gravitational field.
        The atom emits a single photon in a superposition at two possible locations, \textbf{A} and \textbf{B}, with a time difference of $2 T_1$, creating a state consisting of a coherent superposition of two paths (red lines) originating from those locations.
        For simplicity, the photon's recoil on the atom is not displayed in this diagram (see Fig. \ref{fig:atomic_paths} for details).
        The photon's paths are then recombined on a beam splitter and subsequently measured, where again no further measurements are performed on the atom.\newline
        Information about the gravitational field can be obtained from \textbf{a.} the measurement results of the emitted radiation or \textbf{b.} the interfered photon alone.
        The differences in the optical path length of the photons due to different emission events are indicated by the dotted lines.
        Moreover, the radiation/photon that is emitted at \textbf{A} must be delayed (symbolized in the figure by a box with a ``D'') in order to arrive at the beam splitter concurrently with the radiation/photon emitted at \textbf{B} in order to interfere.
    }
    \label{fig:overview}
\end{figure}

Consider the setup depicted in Figure \ref{fig:overview} in which a classical, point-like particle emits radiation in the positive $z$-direction while moving with a velocity $v_0$ along the $x$-direction, within a uniform gravitational field $\vec g = - g \vec e_z$.
Alternatively, $\vec g$ could also be treated as inertial acceleration, which would result in the same procedure.
Radiation emitted at events \textbf{A} and \textbf{B} is directed into an interferometer, where it is recombined at a beam splitter before being measured.
Between the emission events \textbf{A} and \textbf{B}, there is a time difference of $2T_1$.
To compensate for the time difference in emission, a delay must be introduced for the photon emitted at \textbf{A} so that it can simultaneously recombine with the photon emitted at \textbf{B} at the beam splitter.
Such a delay may be realized through a delay line, which consist of an optical fiber or arrangements of mirrors, or a quantum memory\cite{afzelius_quantum_2015}, which can store a photon for a certain time.
Due to their comparatively long coherence and, thus, storage time, quantum memories are especially well-suited for our purposes.
In simple terms, a quantum memory can be realized by using, for instance, an atom with a $\Lambda$ level structure.
An incoming photon can be stored by simultaneously applying a control pulse to the quantum memory, and it can be extracted by applying the control pulse a second time.
The maximum duration a photon state can be stored in a quantum memory without information loss is determined by the coherence time of the atom.%
Quantum memories with coherence times exceeding $1\,$s have already been demonstrated\cite{rancic2017Cohere}.

The time difference in emission results in a path difference traveled by the atom of $2 g T_1^2$ downwards and $2 v_0 T_1$ in the $x$-direction.
This path difference of the body results in an additional path for the radiation, denoted as $\delta$.
Consequently, a phase difference of $\Delta \Phi = 2 \pi \frac{\delta}{\lambda} = \delta k$ arises, where $\lambda$ represents the wavelength and $k$ denotes the wave number of the radiation.
Taking the difference between the intensities $I^\pm$ recorded by the photodetectors $\mathbf{P^-}$ and $\mathbf{P^+}$ and averaging over time yields
\begin{align}
    I^+ - I^- \propto \sin\left( - 2 k v_0 T_1 + 2 k g T_1^2 + \Delta \Phi_\text{c} \right),\label{eq:classical_result}
\end{align}
where $\Delta \Phi_\text{c}$ contains the phases due to the mirror, beam splitter and the photon delay mechanism.
Here we have neglected all contributions of the order of magnitude $\mathcal{O}(v/c)$ and smaller.
A more detailed derivation of this result can be found in the appendix.

Instead of a classical body we now consider a single atom with mass $M$ emitting a single photon.
Like before, the interference of the emitted photon at \textbf{A} and \textbf{B} is measured.
The quantum state consists of photonic states and internal and center of mass degrees of freedom of the atom, which not only form a superposition but rather an entangled state.
Quantum phases accumulated by the photon or atom are common phases of this entire entangled state.
The idea of this setup is, that the quantum phase accumulated by the atom, which includes a gravitationally induced phase, is conveyed to the photonic states by recombining the atom's superimposed paths.

\begin{figure}[b]
    \begin{center}
        \includegraphics[width=0.8\columnwidth]{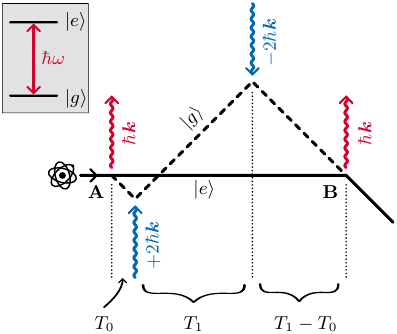}
    \end{center}
    \caption{
        Depicted are the possible paths of the initially excited atom in the atom's freely falling frame.
        The undisturbed path of the atom is illustrated as a solid black line.
        When spontaneously emitting a photon (red), the atom receives a recoil momentum $- \hbar \vec k$.
        If the atom emits the photon at location \textbf{A}, it changes its internal degree of freedom to the $\ket{g}$ state and alters its path, following the dashed black line.
        The dashed path of the atom is redirected twice by laser pulses (blue), which change the momentum by $\pm 2 \hbar \vec k$ without changing the internal degree of freedom \cite{gieseDoubleBraggDiffraction2013}, ultimately recombining the possible paths of the atom at location \textbf{B}.
        Each of these processes induces phase changes of the composed atom-photon system, which are considered in detail in the main text.
        The phase term, including the homogeneous gravitational acceleration $g$, depends on the timing of the blue redirection pulses and can be inferred from the photon measurements when the pulse times are varied.
    }
    \label{fig:atomic_paths}
\end{figure}

The structure of these states is illustrated in Figure \ref{fig:atomic_paths}.
Consider the atom starting out with momentum $\hbar \vec K$ described by the center of mass state $\ket{\vec K}$ in an internally excited state $\ket{e}$ combined with no photons, which is described by $\ket{0}$.
When combined, these states form a product state $\ket{\vec K} \otimes \ket{e} \otimes \ket{0} \eqcolon \ket{\vec K, e; 0}$.
If the atom emits a photon at \textbf{A}, it does so by relaxing from the excited state to the ground state $\ket{g}$.
The energy of the emitted photon is equal to the energy difference between the excited state and the ground state minus the change in the kinetic energy of the atom due to the recoil.
The portion of the momentum of the atom that the photon takes is equal to $\hbar \vec k$, resulting in the product state $\ket{\vec K - \vec k} \otimes \ket{g} \otimes \ket{a_{\vec{k}}} \eqcolon \ket{\vec K - \vec k, g; a_{\vec{k}}}$, where $\ket{a_{\vec{k}}}$ describes a photon emitted at \textbf{A} with momentum $\vec{k}$.
Here we assume the photon emission to be instantaneous, that is, the emission duration is very short compared to the duration of the experiment.
We further assume that the photon is emitted directly into the interferometer.
This situation can be achieved by post-selection or by manipulating the emission process.
For example, a third, non-emitting state of the atom can be utilized, in which the atom is pumped in between the emitting locations.
Another approach is to exploit the Purcell effect by placing cavities at the locations where the photons are to be emitted.
A more detailed discussion of these approaches, together with further ideas for a more rigorous description of the processes shown here, can be found in the section on experimental aspects later in the article.

After the atom has passed \textbf{A}, we assume a 50/50 superposition of the initial state and the state after emission (more general emission amplitudes are discussed later)
\begin{align}
    \ket{\Psi_\text{A}} = \frac{1}{\sqrt{2}} \left( \int \dd^3 k f(\vec k) \ket{\vec K - \vec k, g; a_{\vec{k}}} + \ket{\vec K, e; 0} \right),
\end{align}
where $f(\vec k)$ describes the spectral and directional distribution of the photon.
As one can easily verify, this state is entangled between the internal and center of mass degrees of freedom of the atom and the photonic state.

Next, two laser pulses (blue in Figure \ref{fig:atomic_paths}) are applied that only affect the atom if it has emitted a photon at \textbf{A}.
As a result, this part of the trajectory of the atom is redirected so that it coincides with the atom's trajectory if it propagates freely to point \textbf{B} \footnote{Ideas on how this could be implemented in practice are discussed in the discussions.}.
The emission process at \textbf{B} is assumed to be analogous to the emission process at \textbf{A} resulting in $\int \dd^3 k f(\vec k) \ket{\vec K - \vec k, g; b_{\vec{k}}}$, which resembles the state of emission at \textbf{A}, but with a photonic state $\ket{b_{\vec{k}}}$ instead of $\ket{a_{\vec{k}}}$.
$\ket{b_{\vec{k}}}$ also describes a photon with a sharply defined wave vector $\vec k$, but emitted at \textbf{B} and thus propagating along a different path.

Both parts of the atom's wave function superimpose at \textbf{B} with the same momentum and internal state, but since they arrived via different paths, they have accumulated different quantum phases.
These phases, along with phase contributions from the photons (e.g., due to reflection on the mirror), can be combined into $\Phi^{\text A}$ resp. $\Phi^{\text B}$, which, in general, depend on the photon's wave vector.
A detailed breakdown of $\Phi^{\text{A/B}}$ follows in the next section.
The resulting state after recombination at \textbf{B} can be expressed as
\begin{align}
\begin{split}\label{eq:Psi}
    \ket{\Psi_\text{B}} = \frac{1}{\sqrt{2}} \int \dd^3 k f(\vec k) \text e^{\img \Phi^{\text A}(\vec k)} \ket{\vec K - \vec k} \otimes \ket{g}\\
    \otimes \left( \ket{a_{\vec{k}}} + \text e^{\img (\Phi^{\text B}(\vec k) - \Phi^{\text A}(\vec k))} \ket{b_{\vec{k}}} \right),
\end{split}
\end{align}
where we have used the previously mentioned \textit{path identity}, i.e., that the internal and center of mass degrees of freedom of the atom are identical and can be factored out.
In the following, the dependence of $\Phi^{\text{A/B}}$ on $\vec k$ is not explicitly displayed.
The final state as it arrives at the photodetectors $\mathbf{P^-}$ and $\mathbf{P^+}$ can be calculated by applying the unitary beam splitter transformation
\begin{align}
    \ket{a_{\vec{k}}} \rightarrow& \frac{1}{\sqrt{2}} (\ket{P^-_{\vec{k}}} + \img \ket{P^+_{\vec{k}}}),&
    \ket{b_{\vec{k}}} \rightarrow& \frac{1}{\sqrt{2}} (\img \ket{P^-_{\vec{k}}} + \ket{P^+_{\vec{k}}}).\label{eq:beamsplitter_trafo}
\end{align}
We define the density matrix of the entangled atom-photon state as $\rho = \ket{\Psi_\text{B}} \bra{\Psi_\text{B}}$ and trace out the atom's internal and center of mass degrees of freedom.
Assuming that the width of either the atom's wave packet or the photon emission is much smaller than the other width, and that the total phase does not vary significantly on the scale of $\Delta k$ in the case of narrow photon emission spectra, the reduced density matrix of the photon is given by
\begin{align}
    \rho_{\text{ph}} =& \int \dd^3 K \left( \bra{\vec{K}} \otimes \bra{g} \right) \rho \left( \ket{g} \otimes \ket{\vec{K}} \right)\\
\begin{split}
        =& \frac{1}{2} \int \dd^3 k |f(\vec k)|^2 \bigg[\\
            & \left( 1 + \sin(\Delta \Phi) \right) \ket{P^+_{\vec{k}}} \bra{P^+_{\vec{k}}}
            + \cos(\Delta \Phi) \ket{P^+_{\vec{k}}} \bra{P^-_{\vec{k}}}\\
            +& \cos(\Delta \Phi) \ket{P^-_{\vec{k}}} \bra{P^+_{\vec{k}}}
            + \left( 1 - \sin(\Delta \Phi) \right) \ket{P^-_{\vec{k}}} \bra{P^-_{\vec{k}}}
        \bigg],
\end{split}\label{eq:photon_density_matrix}
\end{align}
where we have introduced the difference of the phases $\Delta \Phi \coloneq \Phi_\text B - \Phi_\text A$.
Detailed discussions on how to obtain this density matrix can be found in the appendix.
The frequency-independent probabilities, $\mathcal{P}^+$ and $\mathcal{P}^-$, of measuring a photon at $\mathbf{P^+}$ and $\mathbf{P^-}$, respectively, can now be obtained from the diagonal elements of $\rho_\text{ph}$,
\begin{align}
\begin{split}
    \mathcal{P}^\pm \coloneq& \int \dd^3k\,\braket{P^\pm_{\vec{k}} | \rho_\text{ph} | P^\pm_{\vec{k}}}\\
    =& \frac{1}{2} \left( 1 \pm \int \dd^3 k |f(\vec k)|^2 \sin(\Delta \Phi) \right),
\end{split}
\end{align}
where we used that $\int \dd^3k f(\vec{k})=1$.
The detection probabilities can also be combined into a single value as $\mathcal{P}^+ - \mathcal{P}^- = \int \dd^3 k |f(\vec k)|^2 \sin(\Delta \Phi)$.
This result is valid under the assumption that the atom is in a state of equal superposition for photon emission at points \textbf{A} and \textbf{B}.
In order to also cover a more general case with arbitrary emission probabilities, an emission amplitude $|\alpha|$ is assumed for the emission at \textbf{A} and an emission amplitude $|\beta|$ for the emission at \textbf{B}.
This modifies the result to
\begin{align}
     \mathcal{P}^+ - \mathcal{P}^- = 2 |\alpha| |\beta| \int \dd^3 k |f(\vec k)|^2 \sin(\Delta \Phi),\label{eq:diff_measurement}
\end{align}
with $|\alpha|^2 + |\beta|^2 = 1$, where the equality holds in the case that the atoms always emit a photon at either \textbf{A} or \textbf{B}.
Importantly, $\Delta \Phi$ depends on the gravitational potential difference which can thus be inferred from measurements of (\ref{eq:diff_measurement}).

In order to ensure that the influence of the gravitational acceleration $\vec g$ is not averaged out we want to emphasize at this point that the width of the spectral distribution of the photon $f(\vec k)$ must be small in comparison to $(g T_1^2)^{-1}$.
For convenience the $|f(\vec k)|^2$ is from now on assumed to be sufficiently narrow that the integral can be omitted and the phase terms are evaluated at a mean $\vec k$ instead.
Given that $k \propto E$, the assumption that the spectral distribution of the photon is narrow implies that the energy uncertainty of the emission process is small.
With regard to the energy-time uncertainty principle, this is in contrast to the assumption, that emission duration (uncertainty) is also short.
However, it is possible to briefly estimate that both assumptions can be satisfied simultaneously.
Doing so leads to simple bounds on the experimental parameters, which are that $\frac{\Delta t}{T_1}$ has to be much larger than $v_0 / 2 \pi c$, $(8 \pi \omega_\text{C} T_1)^{-1/2}$ and $3 g T_1 / 4\pi c$, with the Compton frequency $\omega_\text C = M c^2 / \hbar$, which usually assumes values between $10^{24}\,s^{-1}$ and $10^{27}\,s^{-1}$.

\section{\label{sec:theory}Derivation of the phase}

In the following, $\Phi^{\text{A/B}}$ will be calculated.
The total interferometer phases as used in (\ref{eq:diff_measurement}) consist of four components
\begin{align}
    \Phi^{{\text{A/B}}} = \Phi^{{\text{A/B}}}_\text{at} + \Phi^{{\text{A/B}}}_\text{int} + \Phi^{{\text{A/B}}}_\text{ph} + \Phi^{{\text{A/B}}}_\text{c}
\end{align}
with the atom's internal and center of mass degrees of freedom phase $\Phi^{{\text{A/B}}}_\text{at}$ and $\Phi^{{\text{A/B}}}_\text{int}$, respectively, the photonic phase $\Phi^{{\text{A/B}}}_\text{ph}$ and the constant phase contributions $\Phi^{{\text{A/B}}}_\text{c}$.
This applies to both, the part of the atomic wave function emitting a photon at \textbf{A} and the part emitting at \textbf{B}.
Since only the differences of these phases are of interest, it is sufficient to calculate $\Delta \Phi_i \coloneq \Phi^{\text B}_i - \Phi^{\text A}_i$ with $i \in \{\text{at}, \text{int}, \text{ph}, \text{const}\}$.
In this derivation, all contributions of magnitude $\mathcal{O}(v/c)$ and smaller are neglected.

The simplest part of the total phase is given by the internal degrees of freedom of the atoms.
Here we have
\begin{align}
    \Delta \Phi_\text{int} = - 2 \omega T_1,\label{eq:internal_phase}
\end{align}
since during the time in the interferometer, the two different parts of the atomic wave function are continuously in different energy levels.

For the photonic phase component, the optical path difference between photons emitted at \textbf{A} and photons emitted at \textbf{B} must be considered and is given by $\Delta \Phi_\text{ph} = k \delta$, with the aforementioned path difference $\delta$.
As can be seen in Figure \ref{fig:overview}, a photon emitted at \textbf{A} has an additional travel distance of $2 v_0 T_1$, with the initial velocity in $x$-direction $v_0$.
Together with the photon's temporal evolution, this results in
\begin{align}
    \Delta \Phi_\text{ph} = 2 \omega T_1 - 2 k v_0 T_1.\label{eq:photonic_phase}
\end{align}
Note that a photon emitted at \textbf{B} must travel the additional distance $2 g T_1^2$, which will enter through the spontaneous emission potential in (\ref{eq:pot_se}).

Until now, all derived phase terms had a classical counterpart.
This changes when considering the center of mass degrees of freedom of the atom quantum mechanically.
It is here where the similarity to Quantum Imaging occurs, with the atom corresponding to the idler photon shown in \cite{lemosQuantumImagingUndetected2014}, also establishing what is called \textit{path identity} in later articles.
The derivation of the atom phase is performed similarly as described in \cite{schleichRepresentationfreeDescriptionKasevich2013} by calculating
\begin{align}
    \Phi^{\text{A/B}}_\text{at} =& \frac{1}{\hbar} \int_0^{T} \mathcal L^{\text{A/B}} \dd t\label{eq:atomic_phase},
\end{align}
where $T$ is the travel time of the atom between the two emission events, i.e. \textbf{A} and \textbf{B} in Figure \ref{fig:atomic_paths}.
$\mathcal L^{\text B}$ represents the classical Lagrangian of atoms following the solid line and $\mathcal L^{\text A}$ the classical Lagrangian of atoms following the dashed line.
Accordingly, for this setup, we have $T = 2 T_1$.
The integrals are carried out along the atom's classical path.
While it may seem natural to perform the calculations in the laboratory frame, it turns out to be easier to do so in the freely falling frame of the atom, since the gravitational acceleration then occurs exclusively in the laser pulse terms.
The Lagrangians expressed in such a frame can be written as
\begin{align}
    \mathcal L^{\text A} =& \frac{M}{2} \left[ \dot x^2 + \dot z^2 \right] - V^{\text A}_\text{SE} - V^{\text A}_\text{laser},\label{eq:lagrangian_primed}\\
    \mathcal L^{\text B} =& \frac{M}{2} \left[ \dot x^2 + \dot z^2 \right] - V^{\text B}_\text{SE},\label{eq:lagrangian}
\end{align}
where $\dot z$ represents the atom's velocity in the direction (anti-)parallel to gravity, and $\dot x$ the velocity in the atom's initial direction perpendicular to $\vec g$.
$V^{\text{A/B}}_\text{SE}$ denotes the potentials due to the recoil induced by the spontaneous emission of a photon at either \textbf{A} or \textbf{B} and $V^{\text A}_\text{laser}$ represents the potential due to the laser kicks, which exclusively applies to the part of the atomic wave function emitting at \textbf{A}.
The effective potential due to the laser light is given by \cite{schleichRepresentationfreeDescriptionKasevich2013}
\begin{align}
\begin{split}
    V^{\text A}_\text{laser} = - 2& \hbar k \left( z - \frac{1}{2} g t^2 \right)\\
    &\times \left( \delta(t - T_0) - \delta(t - T_0 - T_1) \right),
\end{split}\label{eq:V_laser_phase}
\end{align}
where it is assumed that $\vec k \parallel \vec e_z$.
Here, the term $- \frac{1}{2} g t^2$ arises from the fact that, in the freely falling frame of the atom, the laser is accelerated by $g$.
Similarly, the potential due to the spontaneous emission of photons and the associated recoil is introduced as \footnote{A derivation of this potential can be found in the appendix.}
\begin{align}
\begin{split}
    V^{\text A}_\text{SE} =& + \hbar k \left( z - \frac{1}{2} g t^2 \right) \delta(t),\\
    V^{\text B}_\text{SE} =& + \hbar k \left( z - \frac{1}{2} g t^2 \right) \delta(t - 2 T_1),\label{eq:pot_se}
\end{split}
\end{align}
with $k$ being the wave number of the photon and we have again replaced $z \to z - \frac{1}{2} g t^2$ to account for the freely falling frame.
The latter results in an additional phase contribution of $+ 2 k g T_1^2$, which is exactly the value we omitted in (\ref{eq:photonic_phase}).
Note that although the $\delta$ distributions in (\ref{eq:pot_se}) are centered on the edges of the integral in (\ref{eq:atomic_phase}), they are considered to lie completely within the integration interval.
This is a reasonable assumption since we could extend the time integral on both sides without altering the result.

Now in order to calculate the integrals in (\ref{eq:atomic_phase}), we first need the classical paths for the atom which we can readily obtain by plugging (\ref{eq:lagrangian}) and (\ref{eq:lagrangian_primed}) into the Euler-Lagrange equation.
Since the result depends only on relative velocities and not on the absolute choice of a coordinate system, we can, without loss of generality, set for this calculation $x(0) = \dot x(0) = z(0) = \dot z(0) = 0$.
Afterwards, evaluating (\ref{eq:atomic_phase}) is straight forward and results in a phase difference of
\begin{align}
    \Delta \Phi_\text{at} =& \frac{\hbar k^2}{M} T_1 - k g T_1 (2 T_0 - T_1)\label{eq:phase_diff}
\end{align}
Here $\frac{\hbar k^2}{M} T_1$ represents the kinetic energy due to the photonic recoil.
The important term is $- k g T_1 (2 T_0 - T_1)$, which is proportional to the gravitational acceleration.

It is convenient to gather all unknown but constant phase contributions due to the photon emission processes, the interaction of the laser pulses with the atom, as well as numerous further sources of constant phase shifts due to optical elements, such as mirrors, beam splitters, lenses, etc. into one term resulting in the constant phase difference $\Delta\Phi_\text{c}$.

By substituting \eqref{eq:internal_phase}, \eqref{eq:photonic_phase} and \eqref{eq:phase_diff} into \eqref{eq:diff_measurement}, we obtain
\begin{align}
\begin{split}
    \mathcal{P}^+ - \mathcal{P}^- = 2 |\alpha| |\beta| \sin\bigg( - 2 k v_0 T_1\\
    + \frac{\hbar k^2}{M} T_1 - k g T_1 (2 T_0 - T_1) + \Delta \Phi_\text{c} \bigg).
\end{split}\label{eq:result}
\end{align}
Since the $\Delta \Phi_\text{c}$ term in this result is unknown, a single absolute measurement of $\mathcal{P}^+ - \mathcal{P}^-$ provides no information about the gravitational acceleration.
One approach to obtain information on $\vec g$ would be to vary the delay $T_0$ in several measurements while keeping the total duration $2 T_1$ constant.
In this way, the part of the atomic wave function that has emitted a photon at point \textbf{A} could be deflected, providing the desired information.

To achieve maximum contrast, the coefficient $|\alpha| |\beta|$ in (\ref{eq:result}) must be maximized under the additional condition $|\alpha|^2 + |\beta|^2 = 1$.
As briefly mentioned before, this is the case for $|\alpha| = |\beta| = \frac{1}{\sqrt{2}}$, i.e., when the emission probability for a photon is $50\,\%$ each for both emission events.

To estimate the sensitivity of the setup we are proposing here, we calculate the Cramér–Rao bound for our scheme with respect to $g$.
The Cramér–Rao bound, which is the reciprocal of the Fisher information, provides a lower bound on the uncertainty of measuring a quantity.
In the case mentioned already briefly before, if the emission width of the photon is regarded to be narrow, the right hand side of \eqref{eq:photon_density_matrix} can be expressed by a pure state
\begin{align}
    \ket{\Psi_\text{ph}} =& \frac{1}{2} \left[ \left( 1 + \img \text e^{\img \Delta \Phi} \right) \ket{P^-_{\vec{k}}} + \left( \img + \text e^{\img \Delta \Phi} \right) \ket{P^+_{\vec{k}}} \right].
\end{align}
This allows for the straightforward calculation of the quantum Fisher information\cite{liu_quantum_2010} for $g$ to
\begin{align}
    \mathcal{F} = \left( \diff{\Delta \Phi}{g} \right)^2.\label{eq:quantum_fisher_information}
\end{align}
Using \eqref{eq:quantum_fisher_information} and \eqref{eq:result}, we can calculate the Cramér–Rao bound to be approximately
\begin{align}
    \Delta g \geq \frac{1}{\sqrt{n} k T_1^2},
\end{align}
where $\frac{1}{\sqrt{n}}$ follows from $n$ repeated independent measurements of $g$.
Using a Mach–Zehnder interferometer to measure gravitational acceleration yields the same Cramér–Rao bound.
In this sense, we expect the setup we propose in this article to have comparable sensitivity bound for measuring $g$ to that of standard atom interferometry.

\section{Experimental aspects}

In the derivation presented in the article, several shortcuts have been employed, which here shall be discussed in more detail.

Firstly, it must be taken into account that both the atomic and photonic wave functions are not point like, but have spatial extensions.
Moreover, the recoil of the photon does not have a single associated momentum, but rather a momentum distribution, coupled with a distribution in the time domain.
This leads to further broadening of the atomic wave function.
Care must be taken to ensure that this broadening remains sufficiently small compared to the path deviation of the atom due to the photonic recoil; otherwise, the quantity we want to measure might be diminished.
A viable approach to address this would be to model the wave functions involved microscopically within the framework of the Wigner-Weisskopf theory of spontaneous emission.
This would also help answering the questions raised in the article regarding the phase relation between the two parts of the wave function.

Another assumption made in the derivation was that the photon is emitted only at two specific events on the atom's trajectory and only in one specific direction.
While this condition can be achieved by post-selecting only those photons that fulfill these criteria, doing so would result in a very low data rate and thus be hard to measure.
Two possibly better methods are:
\begin{itemize}
    \item Utilizing an atom in a third, dark state, one that does not relax over the course of the experiment.
    Initially, at point \textbf{A}, the atom is in an equal superposition of the excited and dark states.
    Subsequently, at point \textbf{B}, this process is repeated: the dark state is pumped into the excited state, leading to immediate relaxation and photon emission.
    As with the laser pulses described in this article, these pump pulses would introduce additional gravitationally dependent phase changes to the atom, thus altering the $g$-dependent term in \eqref{eq:result}.
    Also, the direction of emission can be controlled by aligning the atom's dipole moment.
    \item Leveraging the Purcell effect by placing cavities at the positions where the photons should be emitted would not only increase the emission probability at these events but also enhance emission in the desired direction.
\end{itemize}

Finally, the optical part of the interferometer needs to be addressed in more detail.
This would involve, among other things, introducing a delay for the first emitted photon so that it can recombine simultaneously at the beam splitter with the second emitted photon.
This raises the question of whether a gravitational influence on the delay line could affect the results.
As briefly mentioned in the setup section, a reasonable way to create this delay line experimentally is to use quantum memories.
Different gravitational potentials between the quantum memory and a reference (e.g., the recombining beam splitter) result in time dilation, yielding a gravitational-dependent phase shift of the stored photon \cite{barzel2024QuantumMemories}.
However, changing the timing of the first pulse $T_0$ does not affect this phase contribution and thus does not impact the general results derived in this article.%

\section{Conclusions}

In conclusion, we have introduced a method with the potential for high-precision measurement of the gravitational field based on Quantum Imaging.
Since we consider individual atoms in our setup, we benefit from the same advantages as atomic interferometry, such as high reproducibility and the potential for high-precision gravitational measurements.
Moreover, our approach adds a novel twist, as the information of the gravitational field is completely encoded in the state of the emitted photons.
Therefore, measurement of the photon is sufficient, and no access to the atoms is needed, which opens a possible new route for remote sensing.

In our current derivation, we assume a homogeneous gravitational field, thus limiting our measurements to gravitational and inertial acceleration.
Future work could extend this model to include curvature or the multipole moments of the gravitational potential, thereby enhancing its applicability in geodesy.
Although the assumptions and approximations made in this article are reasonable to demonstrate the feasibility of our setup, high-precision measurements and the aforementioned extensions may require more rigorous calculations.
As briefly mentioned throughout the article, this would involve considering explicit distributions of the photon's wave vector and the atom's momentum.
This, in turn, would require examining the fidelity of the path identity and providing a more detailed model of photon measurement.
Addressing this requires a microscopic treatment of the derivation.

When evaluating the potential for high precision measurements, extensions to gravitational spacetime curvature (that is, inhomogeneous gravitational fields) and multipole moments, it may be worthwhile to investigate their gravitational influence on the delay lines in more detail.
An additional, often discussed question is whether there is a fundamental decoherence in quantum systems, which may result from, e.g., space-time fluctuations, and which will also influence the physics of the delay line as well as the atomic system.
Such questions arise in discussing the measurement process or the quantum-to-classical transition. 
Thus, fundamental questions regarding the foundations of quantum mechanics have direct impact on practical applications.%

\section*{Acknowledgments}

We gratefully acknowledge Emanuel Schlake and Jan P. Hackstein for fruitful discussions on the topic.
M.C. acknowledges funding by the CRC TerraQ from the Deutsche Forschungsgemeinschaft (DFG, German Research Foundation) – Project-ID 434617780 – SFB 1464.
D.R. acknowledges funding by the Federal Ministry of Education and Research of Germany in the project “Open6GHub” (grant number: 16KISK016) and support by the Deutsche Forschungsgemeinschaft (DFG, German Research Foundation) under Germany’s Excellence Strategy – EXC-2123 QuantumFrontiers – 390837967.

\section*{Conflict of Interest}
The authors have no conflicts to disclose.

\section*{Data Availability}
Data sharing is not applicable to this article as no new data were created or analyzed in this study.

\appendix

\section{Classical derivation}
\label{sec:classical_derivation}

A derivation of the classical radiation phase, as shown at the beginning of this article, is presented in more detail here.
Since we are only interested in radiation emitted in (positive) $z$-direction, the emitted electrical field can be regarded as
\begin{align}
\vec E(t, \vec r) =& \vec E_0 \cos(k (z - z_0) - \omega t + \varphi),
\end{align}
where $\vec E_0$ is the amplitude of the electrical field perpendicular to $\vec e_z$, $k$ the wave number of the radiation and $\omega$ is the corresponding frequency connected by $\omega = c k$ with $c$ being the velocity of light.
$\varphi$ represents some arbitrary but constant phase offset.
$z_0$ denotes a spatially offset which in our case contains the extra distance the radiation emitted at \textbf{B} has to travel (not only in $z$-direction, but also in $x$-direction) compared to the radiation emitted at \textbf{A}.
Since the radiating body oscillates in phase with the radiation it emits, there is no temporal offset between the radiation emitted at \textbf{A} and \textbf{B}.
Plugging in the values for this case yields
\begin{align}
    \vec E^\text A(t, \vec r) = \vec E_0 \cos(&k z - \omega t + \varphi_\text A)\\
    \vec E^\text B(t, \vec r) = \vec E_0 \cos(&k (z - 2 v_0 T_1 + 2 g T_1^2) - \omega t+ \varphi_\text B)\label{eq:elec_field_B}
\end{align}
Here we assumed two different phase offsets, which account for the different paths the radiation takes.
The detectors measure the intensity of the radiation, which is proportional to the square of the electrical field averaged over time, i.e.,
\begin{align}
    I^{\pm} \propto \lim_{T \to \infty} \frac{1}{T} \int_{0}^T \left( \vec E^{\pm}(t, \vec r) \right)^2 \dd t.
\end{align}
The electrical field in this formula represents a linear combination of $\vec E^\text A$ and $\vec E^\text B$ with the phase of the electrical field which was reflected at a 50:50 beam splitter shifted by $+\pi/2$.
Evaluating this for both detectors and calculating the difference gives (\ref{eq:classical_result}) where the phase difference $\Delta \Phi_\text{c} = \varphi_\text B - \varphi_\text A$ was introduced and coefficients were omitted.

\section{Derivation of the reduced density matrix of the photon}

In \eqref{eq:photon_density_matrix}, we wrote down the reduced density matrix of the photon, from which the probabilities of measuring a photon at the detectors can be obtained.
Here, we explicitly calculate the steps described in the article and discuss two different physical regimes that both yield the same density matrix.
To keep the article's calculation simple, we used $\ket{\vec K}$ to describe the center of mass motion of the atom.
To discuss the different regimes, we write this state as
\begin{align}
    \ket{\vec K} = \int \dd K^3 F(\vec K) \ket{\overline{\vec K}},\label{eq:initial_state_expanded}
\end{align}
where $F(\vec K)$ describes the expansion of the atom's wavefunction in terms of the momentum eigenstates $\ket{\overline{\vec{K}}}$ with $\int \dd^3 K |F(\vec K)|^2 = 1$.
Calculating \eqref{eq:Psi} again using \eqref{eq:initial_state_expanded} yields
\begin{align}
    \ket{\Psi_\text{B}} =& \frac{1}{\sqrt{2}} \int \dd^3 K F(\vec K) \int \dd^3 k f(\vec k) \text e^{\img \Phi^{\text A}(\vec k)} \ket{\overline{\vec K - \vec k}} \otimes \ket{g}\notag\\
    &\otimes \left( \ket{a_{\vec{k}}} + \text e^{\img (\Phi^{\text B}(\vec k) - \Phi^{\text A}(\vec k))} \ket{b_{\vec{k}}} \right).
\end{align}
As described in the main text, we calculate the reduced density matrix of the photon by tracing out the degrees of freedom of the atom from the density matrix of this state, resulting in
\begin{align}
    \rho_\text{ph} =& \int \dd^3 K' \bra{\overline{\vec{K}'}} \otimes \bra{g} \rho \ket{g} \otimes \ket{\overline{\vec{K}'}}\\
    \begin{split}
    =& \frac{1}{2} \int \dd K'^3 \int \dd^3 k \int \dd^3 k' F^*(\vec K' + \vec k') F(\vec K' + \vec k) \\
    & \times f^*(\vec k') f(\vec k) \text e^{\img ( \Phi^{\text A}(\vec k) - \Phi^{\text A}(\vec k') )}\\
    & \times \left( \ket{a_{\vec{k}}} + \text e^{\img \Delta \Phi(\vec k)} \ket{b_{\vec{k}}} \right) \left( \bra{a_{\vec{k}'}} + \text e^{- \img \Delta \Phi(\vec k')} \bra{b_{\vec{k}'}} \right).
    \end{split}\label{eq:photon_density_matrix_appendix}
\end{align}
Both $F(\vec K)$ and $f(\vec k)$ are functions with a single peak that fall off to zero sufficiently far away from the peak.
Without loss of generality, we can assume that the width of $F(\vec K)$ in all directions is equally $\Delta K$ and that the width of $f(\vec k)$ is $\Delta k$.

The following calculations use the approximation
\begin{align}
\begin{split}
    &\int \dd^3 K F^*(\vec K - \vec k) F(\vec K - \vec k') g(\vec K)\\
    &\quad = \int \dd^3 K F^*(\vec K - \vec k) F(\vec K - \vec k') \delta(\vec k - \vec k') g(\vec K)
\end{split}\label{eq:peak_approximation}
\end{align}
(and analogously with $f(\vec k)$), with a function $g(\vec K)$ for which we will specify the conditions later.
This approximation can be justified as follows:
Since $F(\vec K)$ has only one peak, it can be considered zero everywhere except at the peak. Therefore, the product $F^*(\vec K - \vec k) F(\vec K - \vec k')$ will only yield a non-zero value if the arguments are approximately the same.
This statement can be expressed mathematically using a Dirac delta distribution $\delta(\vec K - \vec k - \vec K + \vec k') = \delta(\vec k - \vec k')$.
Additionally, $g(\vec{K})$ may only vary slightly at this point.
This is the case when the other respective peak function, $f(\vec{k})$ or $F(\vec{K})$, is much wider, i.e., when $\Delta K \gg \Delta k$ or $\Delta K \ll \Delta k$.
Moreover, when $\Delta K \gg \Delta k$, the phase may also only vary slightly on the scale of $\Delta k$ due to the $\vec k$ dependence of the phase.

First, we consider the case of $\Delta K \gg \Delta k$.
Applying \eqref{eq:peak_approximation} to \eqref{eq:photon_density_matrix_appendix} yields
\begin{align}
    \begin{split}
    \rho_\text{ph} \approx& \frac{1}{2} \int \dd K'^3 \int \dd^3 k \int \dd^3 k' F^*(\vec K' + \vec k') F(\vec K' + \vec k) \\
    & \times \delta(\vec k - \vec k') f^*(\vec k') f(\vec k) \text e^{\img ( \Phi^{\text A}(\vec k) - \Phi^{\text A}(\vec k') )}\\
    & \times \left( \ket{a_{\vec{k}}} + \text e^{\img \Delta \Phi(\vec k)} \ket{b_{\vec{k}}} \right) \left( \bra{a_{\vec{k}'}} + \text e^{- \img \Delta \Phi(\vec k')} \bra{b_{\vec{k}'}} \right).
    \end{split}\\
    \begin{split}
    =& \frac{1}{2} \int \dd^3 k |f(\vec k)|^2\\
    & \times \left( \ket{a_{\vec{k}}} + \text e^{\img \Delta \Phi(\vec k)} \ket{b_{\vec{k}}} \right) \left( \bra{a_{\vec{k}}} + \text e^{- \img \Delta \Phi(\vec k)} \bra{b_{\vec{k}}} \right),
    \end{split}
\end{align}
where here, $\delta(\vec k - \vec k')$ originates from the product of the $F(\vec K)$ functions.
Finally, applying \eqref{eq:beamsplitter_trafo} results in \eqref{eq:photon_density_matrix}.

Next, we consider the case that $\Delta K \ll \Delta k$.
Again, we apply \eqref{eq:peak_approximation} to \eqref{eq:photon_density_matrix_appendix}, but now we must perform the approximation using $f(\vec k)$:
\begin{align}
\begin{split}
    \rho_\text{ph} \approx& \frac{1}{2} \int \dd K'^3 \int \dd^3 k \int \dd^3 k'\\
    &\ \phantom{\times} F^*(\vec K' + \vec k' + \vec k) F(\vec K' + \vec k) \\
    & \times f^*(\vec k' + \vec k) f(\vec k) \delta(\vec k') \text e^{\img ( \Phi^{\text A}(\vec k) - \Phi^{\text A}(\vec k' + \vec k) )}\\
    & \times \left( \ket{a_{\vec{k}}} + \text e^{\img \Delta \Phi(\vec k)} \ket{b_{\vec{k}}} \right) \\
    & \otimes \left( \bra{a_{\vec{k}' + \vec k}} + \text e^{- \img \Delta \Phi(\vec k' + \vec k)} \bra{b_{\vec{k}' + \vec k}} \right)
\end{split}\\
\begin{split}
    =& \frac{1}{2} \int \dd^3 k |f(\vec k)|^2\\
    & \times \left( \ket{a_{\vec{k}}} + \text e^{\img \Delta \Phi(\vec k)} \ket{b_{\vec{k}}} \right) \left( \bra{a_{\vec k}} + \text e^{- \img \Delta \Phi(\vec k)} \bra{b_{\vec k}} \right).
\end{split}
\end{align}
Here, we moved $\vec k' \to \vec k' + \vec k$ under the $\vec k'$ integral and then applied \eqref{eq:peak_approximation}, giving $\delta(\vec k')$.
Simplifying the resulting expression yields the same result as before.

\section{Derivation of spontaneous emission potential}

We begin our derivation with the state evolution that results from the Wigner-Weisskopf theory of spontaneous emission, which can be found, for example, in \cite{scullyQuantumOptics1997}, and reads
\begin{align}
\begin{split}
    \ket{\Psi(t)} =& \text e^{-\frac{\Gamma}{2} t} \ket{e, 0}\\
    & + \sum_{\vec k} \alpha_{\vec k} \text e^{- \img \vec k \cdot \vec r_0} \left[ \frac{1 - \text e^{\img \Delta_k t - \frac{\Gamma}{2} t}}{- \Delta_k + \img \frac{\Gamma}{2}} \right] \ket{g, 1_{\vec k}}.
\end{split}
\end{align}
Here $\Gamma$ represents the decay constant, $\vec r_0$ the location of the atom, $\Delta_k = \omega - c k$ the detuning and $\alpha_{\vec k} = \frac{\mathscr E_{\vec k} \vec D \cdot \vec \varepsilon_{\vec k}}{\hbar}$ a generally complex factor, with the dipole moment $\vec D$, the polarization direction $\vec \varepsilon_{\vec k}$ and a normalization constant $\mathscr E_{\vec k}$.
In this calculation, it is assumed that the position of the atom is constant, which is justified by the assumption that the emission process is very short.

The term $\frac{1}{- \Delta_k + \img \frac{\Gamma}{2}}$ is characteristic of relaxation processes and describes a Lorenz-distributed spectrum centered at $\Delta_k = 0$ with a width of $\Gamma$.
To keep this derivation simple, we replace the sum over $\vec k$ by the peak value of the Lorenz curve $\Delta_{k_0} = 0$, i.e. $\vec k_0 = \frac{\omega}{c} \vec e_z$.
A similar assumption is made when deriving the Wigner-Weisskopf theory.
Note that the application of this approximation leads to a non-normalized state, as we omit parts of the state.
This will be corrected later.
Here we choose $\vec k_0$ so that it points in the $z$ direction, since these are the photons we are interested in.
Normally, the emission direction is determined by $\alpha_{\vec k}$ or more precisely by $\vec D \cdot \vec \varepsilon_{\vec k}$ which would, e.g. for S-orbitals, emit equally in all directions.
A more detailed discussion of the emission direction can be found in the article.

The resulting state is
\begin{align}
    \ket{\Psi(t)} =& \text e^{-\frac{\Gamma}{2} t} \ket{e, 0} + \alpha_{\vec k_0} \text e^{- \img k_0 z_0} \frac{1 - \text e^{- \frac{\Gamma}{2} t}}{\img \frac{\Gamma}{2}} \ket{g, 1_{\vec {k_0}}}.
\end{align}
We can now divide the complex factor into modulus and phase terms $\alpha_{\vec k_0} \frac{1 - \text e^{- \frac{\Gamma}{2} t}}{\img \frac{\Gamma}{2}} = \beta \text e^{- \img \varphi}$ and replace the real coefficient $\text e^{-\frac{\Gamma}{2} t} = \gamma$ which results in $\ket{\Psi(t)} = \gamma \ket{e, 0} + \beta \text e^{- \img \varphi} \text e^{- \img k_0 z_0} \ket{g, 1_{\vec {k_0}}}$.
As already mentioned we will have $\beta^2 + \gamma^2 \le 1$ due to the approximations made before.
Since the final state must be normalized after spontaneous emission, we can correct this side effect of our previous approximation by setting $\beta = \sqrt{1 - \gamma^2}$.
Finally, we obtain
\begin{align}
    \ket{\Psi(t)} = \gamma \ket{e, 0} + \sqrt{1 - \gamma^2} \text e^{- \img k_0 z_0 - \img \varphi} \ket{g, 1_{\vec {k_0}}}.\label{eq:SE_final_state}
\end{align}
Here, $\gamma = \gamma(t)$ and $\varphi = \varphi(t)$ are time-dependent.
For $\gamma$ we can conclude that $\gamma(0) = 1$ (we also set this also for negative times) and $\lim_{t \to \infty} \gamma(t) = 0$.

Finally, to derive the potential for spontaneous emission, we first assume that the emission is instantaneous, i.e. $\Gamma \to \infty$.
In this case, $\gamma(t)$ becomes a step function $\gamma(t) \to 1 - \Theta(t) \Rightarrow \sqrt{1 - \gamma^2(t)} \to \Theta(t)$, with the Heaviside function $\Theta(t)$.
This allows the action of the potential to be restricted to an infinitesimally short time period represented by the Dirac delta distribution $\delta(t)$, leading to a model of the potential that is
\begin{align}
    V_\text{SE} = \hbar \left[ k z(t) + \varphi(t) \right] \delta(t - T),\label{eq:pot_se_apx}
\end{align}
where $T$ is the time at which the emission takes place.
As in \cite{schleichRepresentationfreeDescriptionKasevich2013}, we can construct a Hamiltonian with this potential $H_\text{SE} = \frac{p^2}{2 M} + V_\text{SE}$ and neglect the kinetic part.
This allows us to calculate the state directly after the emission by applying the time evolution operator to a state directly before the emission
\begin{align}
    \ket{\Psi(T + \delta)} =& \exp\left( - \frac{\img}{\hbar} \int_{T - \delta}^{T + \delta} V_\text{SE} \dd t \right) \ket{\Psi(T - \delta)},\\
    =& \text e^{- \img k z(T) - \img \varphi(T)} \ket{\Psi(T - \delta)}.
\end{align}
A comparison with the last part of (\ref{eq:SE_final_state}) shows that this potential actually corresponds to our earlier description of spontaneous emission.

For the article, we omit the term $\varphi(t)$ that we calculated in (\ref{eq:pot_se_apx}), since it is independent of position and momentum.
Such a term corresponds to an arbitrary potential offset, which vanishes when dealing with classical Euler-Lagrange equations, as we do here.
These phase terms are assumed to be constant throughout the experiment and are collectively reintroduced at a later time together with other constant phase contributions.

\section{Energy-time uncertainty principle}

In (\ref{eq:diff_measurement}), we assumed that the wave number spectrum of the photon is narrow compared to the inverse of the distance traveled by the atom.
Otherwise, the integral would remain in (\ref{eq:result}).
If the spectrum width, denoted as $\Delta k$, were large, the dependence on $g$ would average out.
This implies that the terms within the sine function in (\ref{eq:result}), with $k$ replaced by $\Delta k$, must be small compared to $2 \pi$.
Consequently, we obtain three inequalities: $2 \Delta k v_0 T_1 \ll 2 \pi$, $\frac{\hbar \Delta k^2}{M} T_1 \ll 2 \pi$, and $\Delta k g T_1^2 \ll 2 \pi$, where we have focused on the leading term in the last inequality.
Furthermore, the energy-time uncertainty principle holds, which states $c \Delta k \Delta t \geq \frac{1}{2}$, of which we use the lower bound.
Combining these inequalities results in the following conditions:
\begin{align}
    \frac{\Delta t}{T_1} \gg& \frac{v_0}{2 \pi c},\label{eq:ineq_velocity}\\
    \frac{\Delta t}{T_1} \gg& \frac{1}{2c} \sqrt{\frac{\hbar}{2 \pi M T_1}} = \frac{1}{\sqrt{8 \pi \omega_\text{C} T_1}},\label{eq:ineq_recoil}\\
    \frac{\Delta t}{T_1} \gg& \frac{g T_1}{4\pi c},\label{eq:ineq_gravity}
\end{align}
where we introduce the Compton frequency $\omega_\text{C}$.
(\ref{eq:ineq_velocity}) and (\ref{eq:ineq_gravity}) suggest that the lower bound for $\Delta t$ is sufficiently small if $v_0 \ll c$ or $g T_1 \ll c$, respectively, indicating a non-relativistic regime, as assumed in this article.
(\ref{eq:ineq_recoil}) results in a small lower bound if $T_1 \gg \omega_\text C^{-1}$, which is fulfilled in consideration of the extremely large Compton frequency $\omega_\text C$ for atoms.

By inserting explicit values \cite{delaguilaBraggGravitygradiometerUsing2018}, we can estimate the order of magnitude of these bounds.
Assuming $^{88}$Sr as the atom species and experimental values of $v_0 = 1\,\text{mm}/\text s$, $T_1 = 0.1\,\text s$, and $g = 9.81\,\text m/\text s^2$, we find that the largest lower bound for the emission time is given by (\ref{eq:ineq_gravity}), approximately $\Delta t \gg 8 \cdot 10^{-11}$\,\text s.
We also required in the article $\Delta t \ll 2 T_1$, which can easily be fulfilled simultaneously with the bounds derived before.

\bibliographystyle{unsrt}
\bibliography{references}

\begin{thebibliography}{10}

\bibitem{kasevichMeasurementGravitationalAcceleration1992}
M.~A. Kasevich and S.~Chu.
\newblock Measurement of the gravitational acceleration of an atom with a
  light-pulse atom interferometer.
\newblock {\em Applied Physics B Photophysics and Laser Chemistry},
  54(5):321--332, May 1992.

\bibitem{petersMeasurementGravitationalAcceleration1999}
A.~Peters, K.~Y. Chung, and S.~Chu.
\newblock Measurement of gravitational acceleration by dropping atoms.
\newblock {\em Nature}, 400(6747):849--852, August 1999.

\bibitem{menoretGravityMeasurements102018}
V.~M{\'e}noret, P.~Vermeulen, N.~Le~Moigne, S.~Bonvalot, P.~Bouyer,
  A.~Landragin, and B.~Desruelle.
\newblock Gravity measurements below 10-9 g with a transportable absolute
  quantum gravimeter.
\newblock {\em Scientific Reports}, 8(1):12300, August 2018.

\bibitem{huDemonstrationUltrahighsensitivityAtominterferometry2013}
Z.-K. Hu, B.-L. Sun, X.-C. Duan, M.-K. Zhou, L.-L. Chen, S.~Zhan, Q.-Z. Zhang,
  and J.~Luo.
\newblock Demonstration of an ultrahigh-sensitivity atom-interferometry
  absolute gravimeter.
\newblock {\em Physical Review A}, 88(4):043610, October 2013.

\bibitem{biedermannTestingGravityColdatom2015}
G.~W. Biedermann, X.~Wu, L.~Deslauriers, S.~Roy, C.~Mahadeswaraswamy, and M.~A.
  Kasevich.
\newblock Testing gravity with cold-atom interferometers.
\newblock {\em Physical Review A}, 91(3):033629, March 2015.

\bibitem{delaguilaBraggGravitygradiometerUsing2018}
R.~P. Del~Aguila, T.~Mazzoni, L.~Hu, L.~Salvi, G.~M. Tino, and N.~Poli.
\newblock Bragg gravity-gradiometer using the {\textsuperscript{1}} {{S}} {
  \textsubscript{0}} -- {\textsuperscript{3}} {{P}} {\textsubscript{1} }
  intercombination transition of {\textsuperscript{88}} {{Sr}}.
\newblock {\em New Journal of Physics}, 20(4):043002, April 2018.

\bibitem{schlippertQuantumTestUniversality2014}
D.~Schlippert, J.~Hartwig, H.~Albers, L.~L. Richardson, C.~Schubert, A.~Roura,
  W.~P. Schleich, W.~Ertmer, and E.~M. Rasel.
\newblock Quantum {{Test}} of the {{Universality}} of {{Free Fall}}.
\newblock {\em Physical Review Letters}, 112(20):203002, May 2014.

\bibitem{taralloTestEinsteinEquivalence2014}
M.~G. Tarallo, T.~Mazzoni, N.~Poli, D.~V. Sutyrin, X.~Zhang, and G.~M. Tino.
\newblock Test of {{Einstein Equivalence Principle}} for 0-{{Spin}} and {{
  Half-Integer-Spin Atoms}}: {{Search}} for {{Spin-Gravity Coupling Effects}}.
\newblock {\em Physical Review Letters}, 113(2):023005, July 2014.

\bibitem{asenbaumAtomInterferometricTestofEquivlancePrinciple2020}
P.~Asenbaum, C.~Overstreet, M.~Kim, J.~Curti, and M.~A. Kasevich.
\newblock Atom-interferometric test of the equivalence principle at the ${10}
  ^{\ensuremath{-}12}$ level.
\newblock {\em Physical Review Letters}, 125:191101, Nov 2020.

\bibitem{mullerAtomInterferometryTestsIsotropy2008}
H.~M{\"u}ller, S.-W. Chiow, S.~Herrmann, S.~Chu, and K.-Y. Chung.
\newblock Atom-{{Interferometry Tests}} of the {{Isotropy}} of {{
  Post-Newtonian Gravity}}.
\newblock {\em Physical Review Letters}, 100(3):031101, January 2008.

\bibitem{dimopoulosAtomicGravitationalWave2008}
S.~Dimopoulos, P.~W. Graham, J.~M. Hogan, M.~A. Kasevich, and S.~Rajendran.
\newblock Atomic gravitational wave interferometric sensor.
\newblock {\em Physical Review D}, 78(12):122002, December 2008.

\bibitem{hongMeasurementSubpicosecondTime1987}
C.~K. Hong, Z.~Y. Ou, and L.~Mandel.
\newblock Measurement of subpicosecond time intervals between two photons by
  interference.
\newblock {\em Physical Review Letters}, 59(18):2044--2046, November 1987.

\bibitem{lopesAtomicHongOu2015}
R.~Lopes, A.~Imanaliev, A.~Aspect, M.~Cheneau, D.~Boiron, and C.~I. Westbrook.
\newblock Atomic {{Hong}}--{{Ou}}--{{Mandel}} experiment.
\newblock {\em Nature}, 520(7545):66--68, April 2015.

\bibitem{zouInducedCoherenceIndistinguishability1991}
X.~Y. Zou, L.~J. Wang, and L.~Mandel.
\newblock Induced coherence and indistinguishability in optical interference.
\newblock {\em Physical Review Letters}, 67(3):318--321, July 1991.

\bibitem{lemosQuantumImagingUndetected2014}
G.~B. Lemos, V.~Borish, G.~D. Cole, S.~Ramelow, R.~Lapkiewicz, and
  A.~Zeilinger.
\newblock Quantum imaging with undetected photons.
\newblock {\em Nature}, 512(7515):409--412, August 2014.

\bibitem{gilabertebassetPerspectivesApplicationsQuantum2019}
M.~Gilaberte~Basset, F.~Setzpfandt, F.~Steinlechner, E.~Beckert, T.~Pertsch,
  and M.~Gr{\"a}fe.
\newblock Perspectives for {{Applications}} of {{Quantum Imaging}}.
\newblock {\em Laser \& Photonics Reviews}, 13(10):1900097, October 2019.

\bibitem{lahiriManyparticleInterferometryEntanglement2018}
M.~Lahiri.
\newblock Many-particle interferometry and entanglement by path identity.
\newblock {\em Physical Review A}, 98(3):033822, September 2018.

\bibitem{hochrainerQuantumIndistinguishabilityPath2022}
A.~Hochrainer, M.~Lahiri, M.~Erhard, M.~Krenn, and A.~Zeilinger.
\newblock Quantum indistinguishability by path identity and with undetected
  photons.
\newblock {\em Reviews of Modern Physics}, 94(2):025007, June 2022.

\bibitem{afzelius_quantum_2015}
M.~Afzelius, N.~Gisin, and H.~De~Riedmatten.
\newblock Quantum memory for photons.
\newblock {\em Physics Today}, 68(12):42--47, December 2015.

\bibitem{rancic2017Cohere}
M.~Rančić, M.~P. Hedges, R.~L. Ahlefeldt, and M.~J. Sellars.
\newblock Coherence time of over a second in a telecom-compatible quantum
  memory storage material.
\newblock {\em Nature Physics}, 14(1):50–54, September 2017.

\bibitem{gieseDoubleBraggDiffraction2013}
E.~Giese, A.~Roura, G.~Tackmann, E.~M. Rasel, and W.~P. Schleich.
\newblock Double {{Bragg}} diffraction: {{A}} tool for atom optics.
\newblock {\em Physical Review A}, 88(5):053608, November 2013.

\bibitem{Note1}
Ideas on how this could be implemented in practice are discussed in the
  discussions.

\bibitem{schleichRepresentationfreeDescriptionKasevich2013}
W.~P. Schleich, D.~M. Greenberger, and E.~M. Rasel.
\newblock A representation-free description of the {{Kasevich}}--{{Chu}}
  interferometer: A resolution of the redshift controversy.
\newblock {\em New Journal of Physics}, 15(1):013007, January 2013.

\bibitem{Note2}
A derivation of this potential can be found in the appendix.

\bibitem{liu_quantum_2010}
W.-F. Liu, L.-H. Zhang, and C.-J. Li.
\newblock Quantum {Fisher} {Information} in {Two}-{Qubit} {Pure} {States}.
\newblock {\em International Journal of Theoretical Physics},
  49(10):2463--2475, October 2010.

\bibitem{barzel2024QuantumMemories}
R.~Barzel, M.~Gündoğan, M.~Krutzik, D.~Rätzel, and C.~Lämmerzahl.
\newblock Entanglement dynamics of photon pairs and quantum memories in the
  gravitational field of the earth.
\newblock {\em Quantum}, 8:1273, February 2024.

\bibitem{scullyQuantumOptics1997}
M.~O. Scully and M.~S. Zubairy.
\newblock {\em Quantum {{Optics}}}.
\newblock Cambridge University Press, 1 edition, 1997.

\end{thebibliography}

\end{document}